\documentstyle[12pt]{article}
\normalsize
\def\pp{\hskip 5mm}
\def\bk{{\bf k}}

\def\bq{{\bf q}}
\def\bP{{\bf P}}
\def\bp{{\bf p}}
\def\b0{{\bf 0}}
\def\bi{{\bf i}}
\def\bj{{\bf j}}
\def\eps{\epsilon}
\def\up{\uparrow}
\def\down{\downarrow}

\def\bra{\langle}
\def\ket{\rangle}
 \def\miceq{= \hskip -5mm \rm ^{^{mic}}}
 \def\meseq{= \hskip -5mm \rm ^{^{mes}}}
 \def\maceq{= \hskip -5.5mm \rm ^{^{mac}}}
 \def\micneq{\neq \hskip -5mm \rm ^{^{mic}}}
 \def\mesneq{\neq \hskip -5mm \rm ^{^{mes}}}
 \def\macneq{\neq \hskip -5.5mm \rm ^{^{mac}}}

\def\Re{{\rm Re}}
\def\Im{{\rm Im}}

\def\cO{{\cal O}}
\def\cF{{\cal F}}
\catcode`\@=11
\def\undernos#1{\mathop{\vtop{\ialign{##\crcr
$\hfil\displaystyle{#1}\hfil$\crcr\noalign{\kern3\p@\nointerlineskip}
$\scriptstyle{\bk,\bp \in {\cal F}} $\crcr\noalign{\kern3\p@}}}}\limits}
\catcode`\@=12

\begin{document}
%
%
\title{Two particle correlations and orthogonality catastrophe
in interacting Fermi systems}
\author{Walter Metzner \\
{\em Institut f\"ur Theoretische Physik C, Technische Hochschule Aachen,} \\
{\em D-52056 Aachen, Germany} \\
and \\
Claudio Castellani \\
{\em Dipartimento di Fisica, Universit\`a "La Sapienza", P.le A. Moro 2,} \\
{\em 00185 Roma, Italy}}
\maketitle
%
%
\begin{abstract}
\pp The wavefunction of two fermions, repulsively interacting in
the presence of a Fermi sea, is evaluated in detail. We consider large but
{\sl finite} systems in order to obtain an unambiguous picture of the
two-particle correlations.
As recently pointed out by Anderson, in $d \leq 2$ dimensions the particles
may be correlated even when situated on the Fermi surface.
The "partial exclusion principle" for two particles with opposite spin on
the same Fermi point is discussed, and related to results from the T-matrix
approximation.
Particles on different Fermi points are shown to be uncorrelated in $d>1$.
Using the results for the two-particle correlations we find that the
orthogonality effect induced by adding an extra particle to a (tentative)
two-dimensional Fermi liquid is finite.
\end{abstract}

\vskip 2cm

%
%
\newpage

\vfill \eject

{\bf 1. INTRODUCTION} \par
\medskip
\pp Several years after Anderson \cite{AND1} conjectured the failure of
Fermi liquid theory in two-dimensional interacting electron systems even
at {\sl weak} coupling, the issue is still rather controversial.
Many body perturbation theory says that Fermi liquid theory
breaks down only in 1D, while it seems valid in any higher dimension, at least
at weak coupling.\cite{PER}
However, the wavefunction for two interacting particles in the presence
of a Fermi sea exhibits a peculiar and at first sight alarming feature,
namely a finite phase shift for two particles with opposite spin sitting
on the same point of the Fermi surface.\cite{AND1} A finite phase shift
signals the presence of correlations in the two-particle wave function, which
seems at odds with the existence of independent and stable quasi particles.
This effect is present at arbitrarily weak coupling in 2D. Anderson \cite{AND1}
indeed argued that due to the finite phase shift the insertion
of an extra particle in a 2D Fermi system causes an orthogonality catastrophe,
making the quasi particle weight vanish, and leading to Luttinger instead of
Fermi liquid behavior.
\par
\pp Within conventional many body theory the phase shift is blurred in the
thermodynamic limit by coarse-graining momentum space, and does not seem to
have any drastic consequences. However, the perturbative many body
formalism may be inadequate when singular correlations
in momentum space appear. Taking the thermodynamic limit before solving the
interacting problem is dangerous in this case.
To obtain an unambiguous picture of the correlations associated with the
finite phase shift and its possible consequences,
it is therefore worthwhile to go back to Schr\"odinger's equation, and analyse
the structure of wave functions in large {\sl finite} systems.
\par
\pp In this work we present a detailed analysis of the correlations between
two locally interacting particles in the presence of a Fermi sea, in one,
two and three dimensions, extending earlier studies by Anderson \cite{AND1}
and by Stamp.\cite{STA} In Sec. 2 we provide some basic definitions and
concepts useful for a clear discussion of large finite systems. In Sec. 3
we solve the two-particle Schr\"odinger equation, and discuss in detail how
the wave functions are affected by the interaction. A careful derivation of
Anderson's "partial exclusion principle" \cite{AND1} for two particles on
common Fermi points will be given, and the controversial relation between
"phase shift" and "phase angle" \cite{ERc,ANDc} will be clarified.
Particles on different Fermi points are shown to be uncorrelated in $d>1$.
In Sec. 4 we will show that the orthogonality effect induced
by adding an extra particle on the Fermi surface of a (tentative) Fermi
liquid is finite in any dimension above one. Hence, in higher dimensions
a breakdown of Fermi liquid theory, if any, must be more subtle than in 1D.
Finally, in Sec. 5, we will conclude with a few remarks on the possibility
of hitherto undetected non-perturbative phenomena at weak coupling.
\par
\bigskip\smallskip

{\bf 2. LARGE FINITE SYSTEMS} \par
\medskip
\pp Our aim is to take the large system limit only {\sl after} having
understood the effects of interactions. It is therefore necessary to define
all quantities appearing in the course of the calculation for {\sl finite}
systems, and introduce certain distinctions which are usually not made in
the infinite volume limit.
\par
\pp For definiteness we consider a one-band Hubbard model on a d-dimensional
simple cubic lattice with lattice constant one and periodic boundary
conditions. The Hilbert space of
states may be spanned by antisymmetrized products of local single particle
states $c^{\dag}_{\bj\sigma} |0\ket$, where $c^{\dag}_{\bj\sigma}$ creates
a fermion with spin projection $\sigma$ on site $\bj$ and $|0\ket$ is the
vacuum. Alternatively one may construct a basis from states with sharp
momentum $|\bk\sigma\ket = a^{\dag}_{\bk\sigma}|0\ket$, where
$a^{\dag}_{\bk\sigma} = V^{-1/2} \sum_{\bj} e^{i\bk\bj} c^{\dag}_{\bj\sigma}$,
and $V = L^d$ is the number of lattice sites. We assume $L$ to be even. The
momenta $\bk$ are taken from the Brillouin zone ${\cal B} = \lbrace \bk =
(k_1,..,k_d): k_{\nu} = (-L/2+1,-L/2+2,\dots,L/2) {2\pi \over L} \rbrace$,
which forms a lattice with $V$ sites and lattice constant $2\pi/L$.
The Hamiltonian is
$$ H = H_0 + H_I =
   \sum_{{\bf ij},\sigma} t_{\bf ij} c^{\dag}_{\bi\sigma}
   c_{\bj\sigma} \> + \> U \sum_{\bf j} n_{\bj\up} n_{\bj\down}
   \eqno(2.1) $$
where $t_{\bi\bj}$ is a (translation invariant) hopping matrix, $U \geq 0$
a (repulsive) coupling constant, and $n_{\bj\sigma} = c^{\dag}_{\bj\sigma}
c_{\bj\sigma}$. The kinetic part can also be written in diagonal form as
$$ H_0 = \sum_{\bk,\sigma} \eps_{\bk} a^{\dag}_{\bk\sigma} a_{\bk\sigma}
   \eqno(2.2) $$
where $\eps_{\bk}$ is the Fourier transform of $t_{\bi\bj}$.
\par
\pp The $N$-particle ground state of the non-interacting system ($U=0$) is
given by
$$ |\Phi_0^N\ket = \prod_{\sigma} \prod_{\bk \in {\cal F}^N}
   a^{\dag}_{\bk\sigma} |0\ket \eqno(2.3) $$
where the non-interacting N-particle Fermi sea ${\cal F}^N$ contains all
momenta in $\cal B$ with $\eps_{\bk} \leq \mu^N$, and the chemical
potential $\mu^N$ is such that ${\cal F}^N$ contains $N/2$ momenta
($N = N_{\up} + N_{\down}$, assume $N_{\up} = N_{\down}$, i.e. N is even).
In the following we will frequently drop the index $N$. Let
$\>\bar{\cal F} = {\cal B}\backslash{\cal F}$ be the complement of the Fermi
sea in ${\cal B}$.
Vectors $\bk^-_F$ denote momenta in ${\cal F}$ on the border to
$\bar{\cal F}$, while $\bk^+_F$ denote momenta in $\bar{\cal F}$
on the border to ${\cal F}$ (see Fig. 1). The set of all
$\bk^-_F$ and $\bk^+_F$ form the "inner" and "outer" Fermi surface $\partial
{\cal F}^-$ and $\partial {\cal F}^+$, respectively. In the limit $L \to
\infty$ both sets define the same manifold $\partial {\cal F}$, the Fermi
surface. We assume $N < L^d$ and $\eps_{\bk}$ to be such that the Fermi
surface is smooth and convex without nesting.
\par
\smallskip
\pp In the following, when considering very large finite systems or,
more precisely, sequences of larger and larger systems, it will be
important to distinguish various levels of "equality" of momenta:
\par
i) "{\sl microscopic}" equality, $\bk' \miceq \bk$, if both are
precisely on the same site of the k-lattice, \par
ii) "{\sl mesoscopic}" equality, $\bk' \meseq \bk$, if both may be
separated by a finite number of steps on the k-lattice, \par
iii) "{\sl macroscopic}" equality, $\bk' \maceq \bk$, if both may
be separated by an infinite number of discrete steps (as $L \to \infty$), which
is smaller than ${\cal O}(L)$, however, such that the distance between $\bk'$
and $\bk$ shrinks to zero for $L \to \infty$. \par
Properties described for $\bk' \maceq \bk$ or $\bk' \meseq \bk$ will be
understood to hold for
"almost" all such cases (zero measure exceptions allowed);
$\bk' \maceq \bk$ is what is usually implied by writing "$\bk' = \bk \>$"
in many-body theory, when performing calculations directly
in the thermodynamic limit, where "$\bk$" refers actually to (infinitely)
many states, and it is supposed that they need not be distinguished any more,
maintaining only their density $V/(2\pi)^d$ in momentum space as the only
information. On the other hand, the Pauli exclusion principle acts only in
the case of "microscopic" equality, but is fortunately easy to build in
exactly, and is all one needs of k-space fine structure in the non-interacting
system. In general, in an interacting system it is not a priori clear
whether the "internal" structure of a "point" in the continuum of momenta in
the infinite volume limit is really irrelevant. A simple (though admittedly
unphysical) example for an interaction where it {\sl is} relevant would be a
strict exclusion principle for particles with {\sl opposite} spin on the same
(in the microscopic sense) point in k-space. The Hubbard or other short range
interactions are of course smooth in momentum space, but singularities in
k-space might be generated non-perturbatively.
\par
\bigskip\smallskip

{\bf 3. TWO PARTICLE WAVE FUNCTION} \par
\medskip
\pp In this section the wave function for two particles with opposite spin
in the presence of a Fermi sea will be evaluated. As in the Cooper problem,
the Fermi sea will be assumed to be {\sl inert}, i.e. its role is merely to
block momentum space.
Much of the calculation in (A) and (B) follows the analysis of the related
problem of two particles on an empty lattice by Fabrizio, Parola and
Tosatti.\cite{FPT}
\par
\medskip

{\bf A) Schr\"odinger equation:} \par
\smallskip

\pp The wave function for two particles with total momentum $\bP$ (conserved)
is a linear combination
$$ |\Psi\ket =
   {\sum_{\bq}}'\> L(\bq) \mid \bP/2+\bq \up,\bP/2-\bq \down \rangle
   \eqno(3.1) $$
where the prime restricts the momenta $\bP/2 \pm \bq$ to $\bar{\cal F}$.
The amplitudes $L(\bq)$ obey the Schr\"odinger equation
$$ (E - E^0(\bq)) L(\bq) =
   {U \over L^d} {\sum_{\bq'}}'\> L(\bq') =: C   \eqno(3.2) $$
where $E^0(\bq) := \eps_{\bP/2+\bq} + \eps_{\bP/2-\bq}$. For $U > 0$
there are two classes of solutions, a trivial class characterized by
$C = 0$, and a non-trivial one with $C \neq 0$, respectively.
\par
\pp In the former case one has eigenvalues $E = E^0$ where $E^0$ is a
non-interacting eigenvalue, and $L(\bq) \neq 0$ only for $\bq$ such that
$E^0(\bq) = E^0$.
In addition, the amplitudes are restricted by the condition
$\sum_{\bq}' L(\bq) = 0$. For each $d^0$-fold degenerate $E^0$, there are
$d^0-1$ such solutions, where $d^0/2$ are spin-triplet and $d_0/2-1$
spin-singlet (if $d_0 \geq 2$). Usually $d^0$ is at least two, due to the
symmetry $E^0(\bq) = E^0(-\bq)$, an exception being $E^0(\b0)$.
\par
\pp In the latter class, one can solve for $L(\bq)$, and obtains
$$ L(\bq) = {C \over E - \eps_{\bP/2+\bq} - \eps_{\bP/2-\bq}}
   \eqno(3.3) $$
while the eigenvalues $E$ are determined by
$$ {1 \over U} =  {1 \over L^d} {\sum_{\bq}}'\>
   (E - \eps_{\bP/2+\bq} - \eps_{\bP/2-\bq})^{-1} =:
    K_L(\bP,E)  \eqno(3.4) $$
Note that $K_L(\bP,E)$ is a {\sl real} function, which has simple poles at
the non-interacting two-particle levels $E^0(\bq)$.
The normalization $1 = \langle \Psi | \Psi \rangle = {\sum_{\bq}}' \> |
L(\bq) |^2$ of the wave function fixes $C$ as
$$ C^{-2} = {\sum_{\bq}}'\> (E - \eps_{\bP/2+\bq} - \eps_{\bP/2-\bq})^{-2}
   \eqno(3.5) $$
For fixed total momentum $\bP$, the non-interacting two-particle levels
$E^0(\bq) = \eps_{\bP/2+\bq} + \eps_{\bP/2-\bq}$ can be ordered in an
ascending sequence $E_0^0,E_1^0,...,E_M^0$.
In terms of $\lbrace E_{\alpha}^0 \rbrace$, the eigenvalue equation reads
$$ {1 \over U} =
   {1 \over L^d} {\sum_{\alpha = 0}^M} {d_{\alpha}^0 \over E - E_{\alpha}^0}
   \eqno(3.6) $$
where $d^0_{\alpha}$ is the degeneracy of the non-interacting level
$E^0_{\alpha}$.
The right hand side has simple poles in $E_{\alpha}^0$. Hence it is obvious
that the solutions of (3.6) also form an ascending sequence $E_0,E_1,...,E_M$,
where $E_{\alpha}^0 \leq E_{\alpha} < E_{\alpha+1}^0$.
\par
\pp Let $|\Phi_{\bk\bk'}\ket$ denote the non-interacting eigenstate obtained
by forming the symmetric (spin-singlet) linear combination of
$|\bk\!\up \bk'\!\down \ket$ and $|\bk'\!\up \bk\!\down \ket$, and possibly
other states with the same energy $E^0 = \eps_{\bk} + \eps_{\bk'}$.
For each $|\Phi_{\bk\bk'}\ket$ there is a corresponding exact eigenstate
$|\Psi_{\bk\bk'}\ket$ of $H$, related to $|\Phi_{\bk\bk'}\ket$ by continuity
as $U \to 0$.
In the following we will analyse the energy shift and the modification
of these wave functions by the interaction. In particular we will calculate
the overlap of interacting and non-interacting wave functions as a
convenient and easy-to-understand measure for the wave function change,
alternative to the "phase shift", which will also be discussed.
\par
\medskip

{\bf B) State $|\Psi_{\bk\bk}\ket$ and "partial exclusion principle":}
\par
\smallskip
\pp Let us now analyse the state $|\Psi_{\bk\bk}\ket$ in dimensions $d =
1,2,3$. We will determine the energy shift $\delta E = E(U) - E^0$,
the overlap $S_{\bk} := \bra\Phi_{\bk\bk}|\Psi_{\bk\bk}\ket$ in the large
volume limit, and, if this overlap is smaller than one, the "range" of the
interacting wave function in k-space.
Setting $\bP = 2\bk$ and extracting the term with $\bq = 0$,
the eigenvalue equation (3.4) becomes
$$ {1 \over U} \> = \> {1 \over L^d \delta E} +
   {1 \over L^d} {\sum_{\bq \neq 0}}' {1 \over \delta E - \Delta E^0(\bq)}
   \> =: \>
   {1 \over L^d \delta E} + \tilde K_L(2\bk,E)
   \eqno(3.7) $$
where $\delta E := E - 2\eps_{\bk}$ and
$\Delta E^0(\bq) := \eps_{\bk+\bq} + \eps_{\bk-\bq} - 2\eps_{\bk}$.
Note that, for small $\bq$,
$$ \Delta E^0(\bq) \sim \sum_{\nu=1}^d a_{\nu} q_{\nu}^2 \quad , \quad
   q_{\nu} = (2\pi/L)n_{\nu} \quad {\rm where} \> n_{\nu} \> {\rm integer}
   \eqno(3.8) $$
i.e. the smallest $\Delta E^0(\bq)$ are of order ${\cal O}(L^{-2})$.
2D turns out to be a critical dimension here, because the distance from
$E^0(\b0) = 2\eps_{\bk}$ to the next non-interacting levels $E^0(\bq)$ is
of order $L^{-2}$ in any dimension, while the potential energy of
$|\Phi_{\bk\bk}\ket$ (as a trial state) is $U/L^d$.
The overlap $S_{\bk}$ is given by the amplitude $L(\b0) = C/\delta E$,
where the normalization constant $C$ can be written as
$$ C^{-2} = {1 \over (\delta E)^2} +
   {\sum_{\bq \neq 0}}' {1 \over [\delta E - \Delta E^0(\bq)]^2}
   \eqno(3.9) $$
\par
\pp For $E$ above $2\eps_{\bk}$ but below the next non-interacting level
$E^0(\bq)$, one has
$$ \tilde K_L(2\bk,E) \to \cases {
   \cO(1)    & for $\bk \miceq \bk_F^+$  \cr
   \cO(L)    & for $\bk \micneq \bk_F^+$ \cr}
   \quad \hskip 6truemm {\rm in} \> d=1 \eqno(3.10a) $$
$$ \tilde K_L(2\bk,E) \to \cases {
   \cO(1)       & for $\bk \meseq \bk_F^+$ \cr
   \cO(\log L)  & for $\bk \mesneq \bk_F^+$  \cr}
   \quad {\rm in} \> d=2 \eqno(3.10b) $$
$$ \tilde K_L(2\bk,E) \to \cO(1)
   \quad \hskip 37truemm {\rm in} \> d=3 \eqno(3.10c) $$
The exception for $\bk \miceq \bk_F^+$ in 1D is due to the complete
blocking of states close to $|\Phi_{\bk\bk}\ket$ by exclusion from the
Fermi sea $\cF$. In $d > 1$ there is no such complete blocking for
$\bk \miceq \bk_F^+$ due to degrees of freedom parallel to the Fermi
surface. Partial blocking makes $\tilde K_L(2\bk,E)$ finite in
2D for $\bk \meseq \bk_F^+$, while it diverges logarithmically otherwise.
Inserting the asymptotic behavior of $\tilde K_L(2\bk,E)$ into the
eigenvalue equation (3.7), one obtains the energy shifts
$$ \delta E \to \cases {
   \cO(L^{-1})    & for $\bk \miceq \bk_F^+$  \cr
   \cO(L^{-2})    & for $\bk \micneq \bk_F^+$ \cr}
   \quad \hskip 11truemm {\rm in} \> d=1 \eqno(3.11a) $$
$$ \delta E \to \cases {
   \cO(L^{-2})       & for $\bk \meseq \bk_F^+$ \cr
   \cO(1/L^2\log L)  & for $\bk \mesneq \bk_F^+$  \cr}
   \quad {\rm in} \> d=2 \eqno(3.11b) $$
$$ \delta E \to \cO(L^{-3})
   \quad \hskip 41 truemm {\rm in} \> d=3 \eqno(3.11c) $$
Rewriting (3.9) as $(S_{\bk})^{-2} = (\delta E)^2/C^2 = 1 \> + \>
(\delta E)^2 {\sum_{\bq \neq 0}}' [\delta E - \Delta E^0(\bq)]^{-2}$,
one thus obtains the overlap
$$ S_{\bk} \to \cases {
   1 - \cO(L^{-1}) & for $\bk \miceq \bk_F^+$  \cr
   r < 1           & for $\bk \micneq \bk_F^+$ \cr}
   \quad \hskip 12truemm {\rm in} \> d=1 \eqno(3.12a) $$
$$ S_{\bk} \to \cases {
   r < 1                      & for $\bk \meseq \bk_F^+$ \cr
   1-{\cal O}(1/(\log L)^2) & for $\bk \mesneq \bk_F^+$  \cr}
   \quad {\rm in} \> d=2 \eqno(3.12b) $$
$$ S_{\bk} \to 1-{\cal O}(L^{-2})
   \quad \hskip 40truemm {\rm in} \> d=3 \eqno(3.12c) $$
For $\bk \micneq \bk_F^+$ in 1D and $\bk \meseq \bk_F^+$ in 2D,
the behavior of $\tilde K_L(2\bk,E)$ implies energy shifts
$\delta E$ of the order of the level spacing, $L^{-2}$, and thus a finite
reduction of the overlap $S_{\bk}$, for any non-zero interaction $U$.
In 1D, the asymptotic overlap $r$ does not depend on $U$, as long as
$U > 0$, while 2D $r$ is U-dependent and goes continously to one for
$U \to 0$.
For $\bk \mesneq \bk_F^+$ in 2D, $\delta E$ turns out to be of order
$\cO(1/L^2\log L)$, i.e. too small for transferring a finite amplitude
to other states besides $|\Phi_{\bk\bk}\ket$. For the exceptional case
$\bk \miceq \bk_F^+$ in 1D, the energy shift $\delta E$ is of order
$L^{-1}$, but here the next allowed levels are separated by a gap.
The reader may compare with the corresponding results for two particles
on an empty lattice in Ref. \cite{FPT}.
\par
\pp The overlap reduction $S_{\bk} < 1$ implies that two particles with
opposite spin cannot fully occupy the same $\bk$-state, a phenomenon
which Anderson \cite{AND1} refers to as "{\sl partial exclusion
principle}".
To clearly see this effect it was important to take the limit
$L \to \infty$ only after having calculated the overlap for finite
systems at finite $U$.
The wave functions $|\Psi_{\bk\bk}\ket$ are very short-ranged in k-space:
The amplitudes $L(\bq)$ in $|\Psi_{\bk\bk}\ket$ are of order $L^{-2}$ as
soon as $\bq$ differs macroscopically from zero, i.e. for $L \to \infty$
the wave function $|\Psi_{\bk\bk}\ket$ is confined to an infinitesimally
small region in momentum space, and is therefore macroscopically
indistinguishable from the non-interacting state $|\Phi_{\bk\bk}\ket$. In
contrast to the case of Pauli exclusion, the state $|\Phi_{\bk\bk}\ket$
is not (even partially) expelled from the Hilbert space of states. For
$L \to \infty$, there is not even a partial transfer of amplitude to
higher energies: Summing the squared probability amplitudes
$|L_{\alpha}(\b0)|^2$ of states with total momentum $\bP = 2\bk$ and
energies $E_{\alpha}$ in an infinitesimal interval around $2\eps_{\bk}$,
the total occupation probability one is recovered.
\par
\medskip

{\bf C) States $|\Psi_{\bk\bk'}\ket$:}  \par
\smallskip

\pp We now analyse the states $|\Psi_{\bk\bk'}\ket$ for generic momenta
$\bk$ and $\bk'$. The energy shift $\delta E = E-\eps_{\bk}-\eps_{\bk'}$
can be determined by splitting the eigenvalue equation (3.4) as
$$ {1 \over U} = {d^0 \over L^d \delta E} +
   L^{-d} {\sum_{\bq}}'' \>
   {1 \over E - E^0(\bq)} =:
   {d^0 \over L^d \delta E} + \tilde K_L(\bk+\bk',E) \eqno(3.13) $$
and studying the asymptotic behavior of $\tilde K_L(\bk+\bk',E)$ for
large $L$. Here $d^0$ is the degeneracy of the non-interacting level
$E^0 = \eps_{\bk} + \eps_{\bk'}$, and $E^0(\bq) := \eps_{\bk+\bq} +
\eps_{\bk'-\bq}$; the double prime indicates $E^0(\bq) \neq E^0$ in
addition to exclusion of $\bk+\bq$ and $\bk'-\bq$ from the Fermi sea
$\cF$. The overlap $S_{\bk\bk'} = \bra\Phi_{\bk\bk'}|\Psi_{\bk\bk'}\ket$
is given by $\sqrt{d^0} C/\delta E$, where $C$ is obtained from
$$ C^{-2} = {d^0 \over (\delta E)^2} +
    {\sum_{\bq}}'' \>
    {1 \over [E - E^0(\bq)]^2}  \eqno(3.14) $$
Recall that $|\Phi_{\bk\bk'}\ket$ is a symmetric combination of $d^0$
degenerate states (with amplitude $1/\sqrt{d^0}$ for each).
For $L \to \infty$, the right hand side of (3.14) is always dominated
by levels in an infinitesimal interval around $E$. The qualitative
behavior of the sum in (3.14) follows from the mean spacing
$\overline{\Delta E^0}$ of levels around $E$, which is related to
the density of two-particle states
$$ \rho(E) = \lim_{\eta \to 0} \lim_{L \to \infty} L^{-d} \>
   {\sum_{\bq}}' \delta_{\eta}(E-\eps_{\bP/2+\bq} -\eps_{\bP/2-\bq})
   \eqno(3.15) $$
via
$$ \overline{\Delta E^0} = d^0(E)/L^d \rho(E)
   \eqno(3.16) $$
Here $\delta_{\eta}(x)$ is a broadened delta-function of width $\eta$,
and $d^0(E)$ is the level degeneracy as determined by symmetry
(accidental degeneracies possible for certain dispersion relations
have zero measure).
We will now discuss results for the overlap $S_{\bk\bk'}$ in various
distinct cases.
\par
\pp If $\bk$ or $\bk'$ (or both) are macroscopically distant from the
Fermi surface, and in addition $\bk' \macneq \bk$, one has a level
spacing of order $L^{-d}$ around $E^0$, and $\tilde K_L$ is finite for
$E$ between $E^0$ and the next non-interacting level. Hence $\delta E$
is of order $L^{-d}$, as the level spacing, which implies that finite
amplitude is transferred to other levels, i.e. $S_{\bk\bk'} < 1$ in any
dimension in this case.
\par
\pp If $\bk \maceq \bk'$ macroscopically distant from the Fermi surface,
one finds
$$ S_{\bk\bk'} \to \cases{
   r < 1  & in $d = 1$ \cr
   1      & in $d = 2$ \cr
   1      & in $d = 3$ \cr }   \eqno(3.17) $$
as is easily understood by
extending the corresponding results for $\bk \miceq \bk'$ in (3.12).
\par
\pp Let us now consider the important case where both momenta lie
macrospocially on the Fermi surface, i.e. $d(\bk,\partial\cF), \>
d(\bk',\partial\cF) \maceq 0$, where $d(.,.)$ denotes the euclidean
distance between points or sets in k-space. In this case, the overlap
obeys
$$ S_{\bk\bk'} \to \cases {
   r < 1 & for $\bk \maceq \bk'$  \cr
   1     & for $\bk \maceq -\bk'$ \cr}
   \quad \hskip 40truemm {\rm in} \> d=1 \eqno(3.18a) $$
$$ S_{\bk\bk'} \to \cases {
   r < 1 & for $\bk \maceq \bk'$ and
   ${d((\bk+\bk')/2,\partial\cF) \over d(\bk,\bk')}$ finite  \cr
   1     & else  \cr}
   \quad \hskip 3truemm {\rm in} \> d=2 \eqno(3.18b) $$
$$ S_{\bk\bk'} \to \hskip 3.5 truemm 1
   \quad \hskip 76.5truemm {\rm in} \> d=3 \eqno(3.18c) $$
for large systems. In deriving these results, the three cases
$\bk \maceq \bk'$, $\bk \maceq -\bk'$ and $\bk \macneq \pm\bk'$ must be
treated separately.
\par
\pp For $\bk \macneq \pm\bk'$ (possible only in $d>1$, for $\bk$ and
$\bk'$ on $\partial\cF$), the two-particle density of states
$\rho(\bk+\bk',\eps_{\bk}+\eps_{\bk'})$ vanishes, i.e. the levels next
to $\eps_{\bk}+\eps_{\bk'}$ are typically at infinite distance on
scale $L^{-d}$. Hence no amplitude is transferred to other levels, and
thus $S_{\bk\bk'} \to 1$ for $L \to \infty$.
\par
\pp For $\bk \maceq -\bk'$ (Cooper channel) the density of states is
finite, but $\tilde K_L(\bk+\bk',E)$ diverges logarithmically for large
$L$ (and $E$ detached from non-interacting levels), i.e. the
Schr\"o\-dinger equation forces $\delta E$ down to order $1/L^d \log L$,
implying $S_{\bk\bk'} = 1 - \cO(1/(\log L)^2)$ in any dimension.
\par
\pp For $\bk \maceq \bk'$ the density of states is divergent in 1D,
zero in 3D, and has a rather subtle behavior in 2D. Let us discuss only
the most difficult (and important) 2D case in detail.
The general qualitative behavior can be understood by assuming a quadratic
dispersion relation $\eps_{\bk} = \bk^2/2$ for simplicity. Setting
$k_F = 1$, the density of two-particle (or two-hole) states in 2D is
then given by \cite{FHN}
$$ \rho(E) = \cases{
   0      &for $\omega < \omega_0$ \cr
   1/4\pi &for $\omega_0 < \omega < \omega_-$ \cr
   {1 \over 2\pi^2} \sin^{-1}{\omega/P \over \sqrt{\omega-\omega_0}}
          &for $\omega_- < \omega < \omega_+$ \cr
   1/4\pi &for $\omega_+ < \omega$ \cr }   \eqno(3.19) $$
Here $\omega$ is the energy relative to $2\eps_F$, i.e. $\omega = E-1$,
and the various regimes are separated by
$\omega_+ = \omega_+(P) = P + P^2/2$, $\omega_- = \omega_-(P) = -P + P^2/2$
and $\omega_0 = \omega_0(P) = (P/2)^2-1$, respectively, where $P = |\bP|$.
See Fig. 2 for an illustration of the various regimes in the
$(P,\omega)$-plane. Note that here we make use only of the part where
$\omega > 0$, corresponding to two particles, not holes.
This density of states has a simple geometric interpretation: For quadratic
dispersion, equi-energy manifolds for two particles with total momentum $\bP$
are spheres (in 2D circles) with center $\bP/2$ in k-space; in 2D, the density
of all states (irrespective of whether particles are in ${\cal F}$ or $\bar
{\cal F}$) is $1/4\pi$ independent of $\bP$ and $\omega$. The density of two
particle states in $\bar{\cal F}$ in 2D is thus simply $x/4\pi$, where
$x \in [0,1]$ is the fraction of diameters crossing the equi-energy circle
with both ends in $\bar{\cal F}$ (see Fig. 3 for an illustration). The same
holds analogously for two holes. The singular behavior of $\rho(E)$ in the
limit $P \to 2k_F$, $E \to 2\eps_F$ is simply due to the fact that $x$ may
assume any value between zero and one, however close to the Fermi surface the
particles may be. Thus it is clear that generically two particles in $\bk$
and $\bk'$ with $\bk' \maceq \bk$ find other levels within an energetic
distance of order $L^{-2}$, the only exception being the cases where $x = 0$,
corresponding to $d((\bk+\bk')/2,\partial{\cal F})/d(\bk,\bk') \to 0$.
It remains to see under which conditions $\tilde K_L(\bk+\bk',E)$ is
finite. In almost all cases there is an infinite number of levels above and
below $E^0 = \eps_{\bk} + \eps_{\bk'}$ since we have required only macroscopic
equality of $\bk$ and $\bk'$. Hence, for $E = E^0 + \delta E$ detached from
non-interacting levels, we may check the finiteness of the sum
$\tilde K_L(\bk+\bk',E)$ for $L \to \infty$ from the corresponding
principal value integral $\Re K(\bk+\bk',E)$, where
$$K(\bP,E) := \lim_{\eta \to 0} \lim_{L \to \infty} K_L(\bP,E+i\eta)
  \eqno(3.20)$$
Note that it doesn't matter whether we insert $K_L$ or $\tilde K_L$ on
the right hand side.
For $P \geq 2k_F$, $K(\bP,E)$ is just the particle-particle bubble known
in many-body perturbation theory (for $P < 2k_F$, however, $K$ differs from
the bubble since the two-hole contribution is absent in $K$).
In the regimes which are of interest here, i.e. $P \geq 2k_F$ and
$E-2\eps_F > \omega_0$ not large, $\Re K(\bP,E)$ can be taken from earlier
results for the particle-particle bubble in 2D \cite{FHN}, i.e.
$$ \Re K(\bP,E) = \cases{
   -{1 \over 4\pi} \log {4(\omega-\omega_0)(\omega_c-\omega) \over
   [-\omega+\sqrt(\omega_+-\omega)(\omega_--\omega)]^2}
    &for $\omega_0 < \omega < \omega_-$ \cr
   -{1 \over 4\pi} \log {4(\omega_c-\omega) \over P^2}
    &for $\omega_- < \omega < \omega_+$ \cr }
   \eqno(3.21) $$
where $\omega_c$ is an ultraviolet cutoff. Obviously $\Re K(\bP,E)$ does
not have a unique limit for $|\bP| \to 2k_F$, $E \to 2\eps_F$. Generically
$\Re K(\bk+\bk',\eps_{\bk}+\eps_{\bk'})$ is finite in the limit
$\bk,\> \bk' \to \partial\cF$, $\bk' \to \bk$, being divergent only
(logarithmically) if the ratio $d(\bk,\bk')/d((\bk+\bk')/2,\partial\cF)$
goes to zero in the limiting process (note that
$d((\bk+\bk')/2,\partial\cF) = P/2-k_F$ and $d(\bk,\bk') =
2(\omega-\omega_0)^{1/2}$).
\par
\pp Hence, {\sl generically} the overlap $S_{\bk\bk'}$ is reduced for $\bk$
and $\bk'$ macroscopically on the same Fermi point in 2D, exceptions being
the rare cases where the ratio $d((\bk+\bk')/2,\partial\cF)/$ $d(\bk,\bk')$
is either zero or infinite. The geometry of the generic and the two
exceptional cases is shown in Fig. 4.
Viewed as a limiting process where $\bk,\> \bk' \to \partial\cF$,
the phase space for this wave function modification in the forward
scattering channel vanishes with
the same power as the one for Cooper scattering. Judging from Fig. 2 the
effect in the forward scattering channel {\sl looks} weaker because
it corresponds to points in a
quadratically narrowing region in the $(P,\omega)$ plain, for $\omega \to
0$, while the Cooper processes take place in the only linearly narrowing
region $\omega > \omega_+(P)$. However, the figure shows only a section
in the $d+1$ dimensional $(\bP,\omega)$ space, where the Cooper processes
take place in a cone around the $\omega$-axis, while the $2k_F$-processes
live in a quadratically narrowing fissure encircling the $\omega$-axis at
a fixed distance $2k_F$. Hence, in both cases the phase space vanishes
quadratically in the low energy limit in 2D.
\par
\pp The results for $S_{\bk\bk'}$ in $d = 1,2,3$ dimensions are summarized
in table 1. Analogous results hold for two holes instead of two particles.
\par
\begin{table}
\begin{center}
\begin{tabular}{|c|c|c||c|c|} \hline
    & \multicolumn{2}{c||}{$d(\bk,\partial\cF)+d(\bk',\partial\cF) \maceq 0$}
    & \multicolumn{2}{c|}{$d(\bk,\partial\cF)+d(\bk',\partial\cF) \macneq 0$}
    \\ \hline
$d$ & $\bk \maceq \bk'$  &  $\bk \macneq \bk'$  &
      $\bk \maceq \bk'$  &  $\bk \macneq \bk'$  \\ \hline
$1$ & $r^{(a)}$ & $1$ & $r$ & $r$ \\
$2$ & $r^{(b)}$ & $1$ & $1$ & $r$ \\
$3$ & $1$ & $1$ & $1$ & $r$ \\ \hline
\end{tabular}
\end{center}
\caption{Values of the overlap $S_{\bk\bk'}$: $r$ denotes an overlap $<1$;
 (a) the special case $\bk \miceq \bk' \miceq \bk_F^+$ has $S_{\bk\bk'} = 1$,
 (b) $S_{\bk\bk'} < 1$ generically for
 ${d((\bk+\bk')/2,\partial\cF)/d(\bk,\bk')}$
 $=$ finite $\neq 0$ and also for $\bk \miceq \bk' \meseq k_F^+$ and
 $\bk \meseq \bk' \miceq k_F^+$ (including $\bk \miceq \bk' \miceq k_F^+$).}
\end{table}

\medskip

{\bf D) Phase shift versus phase angle:} \par
\smallskip
\pp The phase {\sl shift} $\chi_{\alpha}$ for two interacting particles in a
finite system is defined by \cite{AND1,ER}
$$ \chi_{\alpha} = -\pi \delta E_{\alpha}/\Delta E^0_{\alpha}  \eqno(3.22) $$
where $\delta E_{\alpha} = E_{\alpha} - E^0_{\alpha}$, $\Delta E^0_{\alpha} =
E^0_{\alpha+1} - E^0_{\alpha}$. Recall that $\alpha = 0,1,2,\dots,M$ labels
all the different non-interacting two-particle energies $E_{\alpha}^0$ of two
particles with fixed total momentum $\bP$ in a sequence of monotonously
increasing energies, and the interacting eigenvalues obey
$E_{\alpha}^0 < E_{\alpha} < E_{\alpha+1}^0$ for $\alpha = 0,\dots,M-1$.
The phase shift is a measure for
the modification of the non-interacting wave function $|\Phi_{\alpha}\ket$ by
interactions. It is finite if $\bra\Phi_{\alpha}|\Psi_{\alpha}\ket < 1$ and
zero if $\bra\Phi_{\alpha}|\Psi_{\alpha}\ket = 1$. The term "phase shift"
derives from an expression of the form (3.22) for the phase shifts in a
partial wave decomposition in scattering theory.\cite{GOT}
In $d > 1$, $\chi_{\alpha}$ is a wildly fluctuating function of $\alpha$,
which requires a proper average over many levels in order to obtain
a well defined limiting function for $L \to \infty$.
\par
\pp The phase {\sl angle} $\phi(E)$ is defined \cite{ANDc} by
$$ \Gamma(E) = |\Gamma(E)| \exp[i\phi(E)]
   \eqno(3.23) $$
where $\Gamma(E)$ is the 2-particle scattering vertex for an infinite system,
which is given by
$$ \Gamma(E) = {U \over 1 - U K(E)}  \eqno(3.24) $$
and $K(E)$ is obtained from $K_L(E)$, (3.4), via the limiting procedure
$K(E) = \lim_{\eta \to 0}$ $\lim_{L \to \infty} K_L(E+i\eta)$;
the dependence on the total momentum has not been written here.
Note that our $K(E)$ is slightly different from the particle-particle
bubble in perturbation theory, since in $K_L(E)$ two-hole contributions are
absent.
\par
\pp Phase shift and phase angle are in general different quantities,
even for large $L$, except in $d=1$. In \cite{ER} their equivalence has been
shown with the tacit assumption that $\chi_{\alpha}$ tends to a continuous
function $\chi(E)$ as $L$ increases, which is however not generally the case.
Only in 1D $\chi_{\alpha}$ has a unique limit for $L \to \infty$ and
$E_{\alpha}^0 \to E$, and one can indeed show that
$$ \chi(E) :=
   \lim_{L \to \infty \atop E_{\alpha}^0 \to E} \chi_{\alpha} = \phi(E)
   \eqno(3.25) $$
and the overlap of interacting and non-interacting wave functions is related
to the phase shift by the simple identity
$$ \bra\Phi_{\alpha}|\Psi_{\alpha}\ket  \to  {\sin\chi(E) \over \chi(E)}
   \eqno(3.26) $$
in this case. A derivation for these relations is given in Appendix A.
\par
\pp Phase shifts and phase angles for two particles with fixed total
momentum and variable energy are shown in Fig. 5 for a 1D system and in
Fig. 6 for a 2D system. The phase shifts have been calculated for a Hubbard
model with next neighbor hopping $t=1$ and interaction $U=5$, on a large
but finite lattice.
In 1D the relation (3.25) is seen to be verified, while in 2D the phase
shifts fluctuate around the phase angle.
In 2D one may define a function $\bar\chi(E)$ representing the
mean phase shift obtained by averaging many points in a small energy
interval. For the systems studied here it turned out that this mean phase
shift generically differs from the phase angle (at small energies it is
larger), but $\bar\chi(E)$ behaves qualitatively similar to $\phi(E)$, i.e.
one has
$$ \bar\chi(E) = \alpha(E)\phi(E) \eqno(3.27) $$
where $\alpha(E)$ is a smooth function of order one.
In particular, a finite phase angle implies a finite mean phase shift and
vice versa. Furthermore, following the steps in \cite{ER}, it is easy to
see that there is a general identity relating the phase angle to the
average {\sl energy} shift, namely
$$ \phi(E) = -\pi \overline{\delta E}/\overline{\Delta E^0} \eqno(3.28) $$
where $\overline{\Delta E^0}$ is the mean level spacing.
\par
\pp A controversy in previous studies arose on whether the phase
shift for two particles on the same point of the Fermi surface in 2D is
finite or not, the problem being that the limit $\bP \to 2\bk_F$, $E \to
2\eps_F$ is not unique. It was noticed that the asymptotic phase angle
is finite if the limit is taken in a particular way.\cite{STA,FHN,ER}
However, as we have pointed out above, what looks so special a limit in
the $(P,E)$ plane reflects actually the generic behavior of two particles
with momenta $\bk'$ and $\bk$ in the limit $\bk' \to \bk \to \partial\cF$.
Anderson's \cite{AND1} finite phase shift, which has originally been
calculated in the special case $\bk \miceq \bk'$, is generically present
for $\bk \maceq \bk'$, too.
\par
\pp The detailed behavior of the phase angle $\phi(\eps_{\bk} +
\eps_{\bk'})$ with $\bP = \bk + \bk'$ for two particles in $\bk$ and $\bk'$
in the limit $\bk' \to \bk \to \partial\cF$ is illustrated in Fig. 7.
\par
\medskip

{\bf E) Antibound states:} \par
\smallskip
\pp To complete the presentation of the two-particle problem in a lattice
model (with an upper bound in energy), we now briefly discuss properties
of the "antibound state" \cite{AND1} on top of the two-particle spectrum.
\par
\pp Since $E_{\alpha} < E_{\alpha+1}^0$ for $\alpha \leq M-1$, all levels
but the highest (for a given total momentum $\bP$) are shifted only by a
tiny amount of order $L^{-d}$ or even less in some cases. However
$\delta E_M = E_M - E_M^0$ turns out to be finite if $U$ exceeds a critical
value $U_c$, which depends on density and dimensionality. In $d \leq 2$,
(3.4) implies that $U_c = 0$, since the density of two-particle levels
is finite (in 2D) or divergent (in $d < 2$) at the upper band edge. For
small $U$, $\delta E_M$ is exponentially small in 2D, while it is of order
$U^2$ in 1D. This split state has been called the "antibound
state".\cite{AND1} It is in a sense the mirror image of the bound state in
the Cooper problem with attractive $U$.
\par
\pp The antibound state $|\Psi_M\ket$ is actually a {\sl bound} state in
that it has a finite expectation value for double occupancy of sites in real
space even for $L \to \infty$, i.e.
$$ \bra\Psi_M|H_I|\Psi_M\ket \sim {\cal O}(1)
   \quad {\rm for} \quad U > U_c  \eqno(3.29) $$
This is of course energetically highly unfavorable for repulsive (positive)
$U$, which is why $\delta E_M$ is positive of order one. Since the trace of
$H_I$ in the subspace of two-particle states with fixed total momentum is
always $U$, independent of the basis, equation (3.29) implies that
$$ \sum_{\alpha = 0}^{M-1} \bra\Psi_{\alpha}|H_I|\Psi_{\alpha}\ket < U
   \quad {\rm for} \quad U > U_c  \eqno(3.30) $$
Hence the splitting of an antibound state means that the other states in the
continuum have an overall reduced expectation value for double occupancy,
and $|\Psi_M\ket$ just pays the bill for all of them. In the Hilbert space of
two-particle states spanned by $\lbrace |\Psi_{\alpha}\ket: \alpha = 0,\dots,
M-1 \rbrace$ double occupancy has been partially projected out. In 2D, this
projection is a weak coupling effect (present for any $U>0$), in 3D not.
\par
\pp Ignoring states which are separated by a gap from the low-energy part of
the spectrum, one may say that the presence of a down-spin, say, reduces the
dimensionality of the space of available states for up-spin particles, as
in the presence of statistical interactions \cite{HAL1} between opposite
spins. The significance of the antibound states in the {\sl two}-particle
system for the {\sl many}-particle system is however not yet clear.
Anderson \cite{AND1} suggested that the splitting of the antibound states
implies that the Hubbard model in $d \leq 2$ might have the same low energy
behavior as the tJ-model, where doubly occupied sites are projected out
completely.
A full projection of double occupancy can be implemented by gauge fields,
leading to singular effective interactions which have been argued to
invalidate Fermi liquid theory for the tJ-model (at least at finite
temperatures).\cite{GAU}
Thus, {\sl if} the above (controversial!) arguments were valid, Fermi liquid
theory would break down in the 2D Hubbard model for any $U > U_c = 0$.
\par
\bigskip\smallskip

{\bf 4. ORTHOGONALITY CATASTROPHE} \par
\medskip

\pp Long ago Anderson \cite{AND3} pointed out that a {\sl local} scatterer
in a many fermion system changes the wave function so drastically that its
overlap with the wave function without scatterer is zero in the infinite
volume limit, and related this "orthogonality catastophe" to the observed
singular response of electron systems to a sudden appearance of local
scatterers, such as in the X-ray problem. More recently he proposed to
extend this line of reasoning to the insertion of a quasi particle in an
interacting Fermi system.\cite{AND1}
\par
\pp To understand the argument, it is useful to recall the case of a local
scatterer first. A system of non-interacting (spinless) fermions in the
presence of a local potential on site ${\b0}$ is governed by the Hamiltonian

$$ H = \sum_{\bk} \eps_{\bk} n_{\bk} +
   U L^{-d} \sum_{\bk,\bq} a^{\dag}_{\bk+\bq} a_{\bk}  \eqno(4.1) $$

The ground state of $H$ is a Slater-determinant constructed with single
particle wave functions of the form
$|\Psi_{\bk}\ket = \sum_{\bq} L_{\bk}(\bq) a^{\dag}_{\bk+\bq} |0\ket$
where $L_{\bk}(\bq) = C_{\bk}/(E_{\bk}-\eps_{\bk+\bq})$, the eigenvalue
$E_{\bk}$ is the solution next to $\eps_{\bk}$ of
$U^{-1} = L^{-d} \sum_{\bq} (E - \eps_{\bk+\bq})^{-1}$, and $C_{\bk}$ is
fixed by normalization. Note that $\delta E_{\bk} = E_{\bk} - \eps_{\bk}$
and $C_{\bk}$ are both typically of order $L^{-d}$, corresponding to
the spacing of the non-interacting levels (except for $\bk \maceq \b0$).
The overlap of the Fermi seas with and without scattering potential,
respectively, is given by \cite{AND3}

$$ S := \bra\Phi|\Psi\ket = \undernos{det} (L_{\bk\bp})  \eqno(4.2) $$

where $L_{\bk\bp} := L_{\bk}(\bp-\bk)$. Note that $L_{\bk\bp}$ decays
rapidly as a function of $\eps_{\bp}-\eps_{\bk}$, but the momentum transfer
$\bp-\bk$ may be large. A sufficient condition for the orthogonality
catastrophe $S \to 0$ when $L \to \infty$ is that the sum
$$ s:= \sum_{\bk \in {\cal F}} \sum_{\bp \in \bar{\cal F}} |L_{\bk\bp}|^2
   \eqno(4.3) $$
diverges. Indeed $s$ behaves roughly as $L^{-2d}\sum_{\bk \in {\cal F}}
\sum_{\bp \in \bar{\cal F}} (\eps_{\bk}-\eps_{\bp})^{-2}$,
which is logarithmically infrared divergent
for $L \to \infty$ in any dimension. Note that contributions
come from any $\bq = \bp-\bk$ (not only small ones) across the Fermi surface.
To obtain the orthogonality catastrophe it is important that all (or at least
a finite fraction of) the single particle states $a^{\dag}_{\bk}|0\ket$ with
$\bk \in \partial{\cal F}$ are modified by the scatterer, i.e.
$\bra\Phi_{\bk}|\Psi_{\bk}\ket < 1$ or, what is the same, the phase shift
$\chi_{\bk} := \delta E_{\bk}/\Delta E_{\bk}$ must be finite all over the
Fermi surface ($\Delta E_{\bk}$ being the distance to the level following
$\eps_{\bk}$).
\par
\pp Recently Anderson \cite{AND1} suggested to infer the breakdown of Fermi
liquid theory in two-dimensional interacting electron systems from an
orthogonality catastrophe caused by insertion of an extra particle.
To this end he considers the overlap

$$ Z_{\bk}' = |\bra a^{\dag}_{\bk\sigma} \Psi_0^N |
   \Psi_{\bk\sigma}^{N+1} \ket|^2  \eqno(4.4) $$

for "$\bk = \bk_F$".
Here $\Psi_0^{N}$ is the exact ground state with $N$ particles while
$\Psi_{\bk\sigma}^{N+1}$ is an exact eigenstate of the $N+1$ particle system
with "one quasi particle added", i.e. the state evolving adiabatically from
the non-interacting state $a^{\dag}_{\bk\sigma}|\Phi_0^N \ket$, as the
interaction is switched on. In analogy to the local scatterer problem it is
argued that the finite phase shift inflicted by the extra particle on the
other particles on the Fermi surface will lead to an orthogonality catastrophe,
$Z_{\bk_F}' \to 0$, and consequently the elimination of the quasi particle
peak in the spectral function.
\par
\smallskip
\pp This argument presents various difficulties which we will now discuss.
\par
\smallskip
\pp (i) The macroscopic spectral function $\rho(\bk,\xi)$ is given in terms
of exact eigenstates by
\vskip -1mm
$$ \rho(\bk,\xi) = \lim_{\eta \to 0} \lim_{L \to \infty}
   \sum_n |\bra \Psi_{\bk\sigma,n}^{N+1}|a^{\dag}_{\bk\sigma}|\Psi_0^N \ket|^2
   \delta_{\eta}(\xi-\xi_n)  \eqno(4.5) $$
\vskip -1mm
for positive energies $\xi > 0$; the sum runs over
all (N+1)-particle eigenstates with momentum $\bk$ and spin $\sigma$ (relative
to $|\Psi_0^N \ket$), $\xi_n$ is the excitation energy, and $\delta_{\eta}$ a
broadened $\delta$-function, e.g. $\delta_{\eta}(x) = \eta/\pi(x^2 + \eta^2)$.
Each "point" $(\bk,\xi)$ in $\rho(\bk,\xi)$ involves actually an infinite
number of eigenstates of the interacting system. Hence the vanishing of the
overlap with a {\sl single} eigenstate in the large system limit does not
necessarily affect $\rho(\bk,\xi)$.
In particular, it is easy to see that $Z_{\bk}' \to 0$ for generic
$\bk \maceq \bk_F$ in any dimension, even in a Fermi liquid. By definition,
in a Fermi liquid the spectral function obeys the asymptotic behavior
$$\rho(\bk,\xi) \to Z_{\bk} \delta_{\Gamma_{\bk}}(\xi-E_{\bk})
  \eqno(4.6)$$
for $\xi \to 0$, $\bk \to \partial\cF$, where $E_{\bk}$
is the quasi particle energy, $Z_{\bk}$ a finite renormalization constant,
and $\Gamma_{\bk}$ the width of the quasi particle peak, which must vanish
more rapidly than the quasi particle energy when approaching the Fermi
surface.
In a Fermi liquid, for $\bk \maceq \bk_F$ the width of the quasi particle
peak is zero on scale one, but generically infinite on the scale set by the
level spacing. Hence, $Z_{\bk}'$ is zero in this case.
To have a chance to get a finite $Z_{\bk_F}'$, one must set at least
$\bk \meseq \bk_F^+$, which would however leave the numerical value of
$Z_{\bk_F}'$ completely arbitrary, if finite (e.g., for $\bk$ a hundred
steps on the k-lattice away from $\partial\cF$, $Z_{\bk}'$ is much
smaller than for $\bk \miceq \bk_F^+$. A reasonable unique definition
of a possibly finite $Z_{\bk_F}'$ requires the choice $\bk \miceq \bk_F^+$
in (4.4). We are not aware of a general identity relating $Z_{\bk_F}'$ to
the renormalization factor $Z_{\bk_F}$ in (4.6). A priori $Z_{\bk_F}'$
might still vanish even if $Z_{\bk_F}$ is finite (but not vice versa,
of course).
\par
\smallskip
\pp (ii) In contrast to the local scatterer problem, the extra particle
inserted here has a complicated dynamics, and the overlap (4.4) cannot be
calculated exactly.
Making an estimate in analogy to the case of a local scatterer
added to a Fermi gas amounts to making two (independent) approximations,
which may miss important physics:
The system in the absence of the extra particle is treated as
non-interacting, i.e. the ground state $|\Psi_0^N\ket$ becomes simply a
product of  two non-interacting Fermi seas for up and down spins, i.e.
$|\Psi_0^N\ket \approx |\Phi_0^N\ket$.
Only interactions between the extra particle (with spin down, say) and
particles with opposite spin (i.e. up) are kept. Still the calculation of
$|\Psi_{\bk_F\down}^{N+1}\ket$ poses a many-body problem, due to effective
interactions between up-spins mediated by the extra down-spin: an up-spin
may scatter the down-spin to a new state, which changes its relation to
other up-spins.\cite{KRS}
For a local scatterer (without internal degrees of freedom) this problem
does not occur, because the scatterer remains always in the same state.
Neglecting these induced correlations, too, one may estimate the overlap
$\bra a^{\dag}_{\bk_F\sigma} \Psi_0^N |\Psi_{\bk_F\sigma}^{N+1} \ket$
in analogy to the problem of a local scatterer by approximating
$$ \bra a^{\dag}_{\bk_F\sigma} \Psi_0^N |\Psi_{\bk_F\sigma}^{N+1} \ket
   \> \approx \> \undernos{det} (L_{\bk\bp}) \eqno(4.7) $$
where $L_{\bk\bp} := L_{\bk}(\bp-\bk)$,
and the amplitudes $L_{\bk}(\bq)$ are extracted from the interacting
two-particle wave functions evolving from $|\Phi_{\bk\bk_F}\ket$, i.e.
$$ |\Psi_{\bk\bk_F}\ket = {\sum_{\bq}}' L_{\bk}(\bq)
   |\bk+\bq \up \bk_F-\bq \down \ket  \eqno(4.8) $$
solved for fixed $\bk' \miceq \bk_F^+$ and variable $\bk \in \cF$ in
presence of an inert Fermi sea of down-spins, but no other up-spins.
Here only the down-spin momentum $\bk_F-\bq$ is blocked by exclusion from
$\cF$.
Note that only the up-spins reduce the overlap, since the down spin Fermi
sea is treated as inert. Since the up-spins are not blocked by a
pre-existent Fermi sea, most $|\Psi_{\bk\bk_F}\ket$ will be shifted from the
corresponding non-interacting states $|\Phi_{\bk\bk_F}\ket$. However, most
modifications inside $\cF$ in k-space cancel out when the Fermi sea is
filled, and only shifts leading out of $\cF$, as measured by the sum
$s = \sum_{\bk \in {\cal F}} \sum_{\bp \in \bar{\cal F}} |L_{\bk\bp}|^2$,
are relevant. These latter shifts are of the same order of magnitude as
those considered in Sec. 3, where $|\Psi_{\bk\bk'}\ket$ has been analyzed
for $\bk$ and $\bk'$ on the surface of two already pre-existent Fermi seas.
\par
\smallskip
\pp
iii) In Sec. 3 we have seen that two particles on the Fermi surface may be
scattered within $\bar\cF$ only if they are on the {\sl same} point of the
surface. Analogously, an extra down-particle added to the system in $\bk_F$
in the presence of an inert Fermi sea down-spins may scatter up-particles
out of $\cF$ into $\bar\cF$ only in $\bk_F$ itself, but not on other points
of the surface. Even worse, if $\bk_F \miceq \bk_F^+$, in most cases the
extra particle is not able to scatter {\sl any} up-spin out of $\cF$ (this
is slightly different from the situation in Sec. 3, where both particles
were outside $\cF$ from the start, and thus could always shift parallel
to the Fermi surface in $d>1$).
Comparing this state of affairs with the response to a {\sl local}
scatterer, where a wave function change over the {\sl whole} Fermi surface
led to a logarithmically (only) divergent signal of orthogonality, it is
obvious that here we find no signal at all.
\par
\pp
In summary, a straightforward adaption of the local scatterer calculation
to the problem of inserting a dynamical particle into an interacting
many-body system does not signal an orthogonality catastrophe in 2D.
Stamp,\cite{STA} too, concluded that considering finite systems within
a two-particle scattering approximation does not yield any evidence for an
orthogonality catastrophe. Clearly, approximating the ground state by a
Fermi gas may give qualitatively correct results only if the exact ground
state is a Fermi liquid. Hence, as in perturbation theory, we have only
checked {\sl consistency} of quasi particle behavior as a hypothesis.
We have to recognize that the above check of orthogonality is insufficient,
if the exact ground state is neither a Fermi liquid nor a state obtained
by resumming divergencies showing up in perturbation theory (as in the
one-dimensional Luttinger liquid \cite{SOL,HAL2}).
\par
\smallskip
\pp Note that the phase shift as calculated in Sec. 3 does {\sl not} signal
the orthogonality
catastrophe which is known to occur in a 1D interacting Fermi system upon
adding an extra particle near one of the two Fermi points!
A particle inserted in $\bk \miceq \bk_F^+$ cannot kick out any states near
$\bk_F^+$ itself in 1D (and for $\bk \meseq \bk_F^+$ it can affect only a
finite number). Hence interactions with particles near the {\sl same} Fermi
point do not produce an
orthogonality catastrophe in 1D, whether the phase shift is finite or not.
On the other hand, interactions with particles on the {\sl opposite} Fermi
point do affect infinitely many states, and second order perturbation
theory does indeed indicate an orthogonality catastrophe in this case.
However, the phase shift calculated in Sec. 3 turned out to be zero in this
case, vanishing logarithmically in the large system limit.
This is an artefact of our treating the Fermi surface as inert, not allowing
for particle-hole excitations when calculating two-particle correlations.
In a diagrammatic language, treating the Fermi sea as inert means summing
only ladder diagrams, which is equivalent to introducing a renormalized
coupling whose flow is calculated from the particle-particle
channel only, and therefore seems to scale to zero logarithmically (for
positive bare coupling). In 1D there are however other contributions,
involving particle-hole excitations, which make the $\beta$-function vanish
identically,\cite{SOL,DM} i.e. the renormalized coupling and the exact phase
shift in the many-body system remain finite, and an orthogonality catastrophe
does occur.
\par
\bigskip\smallskip

{\bf 5. CONCLUSION}
\medskip

\pp Two fermions in the presence of a Fermi sea can have interaction
induced correlations even if both particles are situated on the Fermi
surface: In addition to the well-known Cooper pair correlations for
attractive interactions, in low dimensional systems correlations appear
even for purely repulsive interactions, namely i) if both particles sit
on the same point of the Fermi surface (in $d \leq 2$), and ii) if they
sit on opposite points (in $d = 1$). In the former case a solution of
the Schr\"odinger equation for a two-particle wave function in presence
of an {\sl inert} Fermi sea reveals these correlations, while in the
latter case one must allow for particle-hole excitations to obtain the
correct result for a many-body system. Equivalently, in the former
case a properly interpreted T-matrix calculation yields the effect,
while in the latter a complete one-loop renormalization group
calculation of the two-particle vertex is required.
\par
\pp In the full many-body problem we must actually distinguish
correlations between {\sl bare} particles and correlations between low
energy excitations (i.e. {\sl quasi} particles in a Fermi liquid).
Correlations among bare particles are of course always present in
an interacting theory, but are largely absorbed in the wave function
renormalization when passing to an effective theory of the low energy
excitations. The issue here is whether there are correlations between
(tentative) {\sl quasi} particles, surviving at arbitrarily low energy
scale. Of course the above refers to these latter correlations only.
\par
\pp It was important to distinguish various scales of distances on the
k-lattice of momenta, to obtain a clear picture of the rather
singular correlations in k-space, and to relate Anderson's \cite{AND1}
results for the phase shift to results from the T-matrix
approximation.\cite{PER} Anderson's finite phase shift,
calculated for two particles residing on the very same point of the
k-lattice, was seen to be not an artefact of this
special choice, but represents the {\sl generic} behavior in the limit
$\bk,\bk' \to \partial\cF$ with $\bk' \to \bk$. This behavior is in fact
correctly signalled by the corresponding limit of the phase angle of
the scattering vertex calculated in T-matrix approximation.
\par
\pp Two fermions on the Fermi surface of a two-dimensional system repel
each other at very short distances in momentum space. We note that this
"partial exclusion principle" \cite{AND1} is not only "partial", but also
less persistent than genuine statistical correlations such as Pauli
exclusion: The expectation value $\bra n_{\bk_F\up} n_{\bk_F\down} \ket$
will rise when more and more particles are added to the system, and may
come arbitrarily close to one. The amplitude for two particles on the
same Fermi point is only partitioned among different eigenstates with
energies in an infinitesimal interval at $2\eps_F$.
However, the {\sl independence} of quasi particles is obviously spoiled
by these correlations at short distances in k-space. Since quasi particles
on different Fermi points are however uncorrelated in 2D, Landau
parameters involving a smooth angular average may still be well defined.
\par
\pp The orthogonality effect caused by addition of an extra particle
on the Fermi surface of a two-dimensional Fermi gas was shown to be
finite within a crude approximation which takes into account only
two-particle correlations between the extra particle and other particles
in the system. This confirms earlier consistency checks of Fermi liquid
theory performed directly in the infinite volume limit,\cite{FHN,ER}
and a recent study showing the irrelevance of finite size effects in the
T-matrix approximation.\cite{STA}
In contrast to the case of addition of a {\sl local}
scatterer, which modifies the wave functions of other particles over the
whole Fermi surface, a quasi particle in 2D modifies the wave functions
at best on that Fermi point where it is added. In one dimension an
orthogonality catastrophy does occur as a consequence of finite phase
shifts for particles on {\sl opposite} Fermi points.
\par
\pp All well established weak coupling instabilities of the Fermi liquid
are signalled by the renormalization group, evaluated perturbatively to
some low order.\cite{SOL,GAL,FT,SHA}
A recent analysis of the crossover from 1D Luttinger liquid behavior to
2D Fermi liquid behavior as a function of continuous dimensionality within
perturbation theory summed to all orders indicates that at weak coupling
higher orders in perturbation theory do not destroy the Fermi liquid fixed
point in any dimension above one, as long as no Cooper instability sets
in.\cite{CDM}
In addition, recent rigorous results on two-dimensional Fermi systems
seem to indicate that the existence of hitherto unknown weak coupling
instabilities of the Fermi liquid is unlikely.\cite{FT}
These rigorous results are however not yet general enough to be applied to
a system like the Hubbard model.
\par
\pp On the other hand, the rather peculiar change of the two-particle
wavefunction, especially in 2D, could throw doubts on the general validity
of conventional many-body theory itself, even if summed to all orders.
Two particles near the same Fermi point indeed develop singular
correlations in k-space, which are however visible only if the
discrete fine structure of k-space is resolved. These correlations
might not be adequately taken into account when taking the infinite volume
limit before solving the full interacting problem. In a two-dimensional Fermi
liquid ground state, pairs of up- and down-spins near a common Fermi point
would seem quite unstable objects: they have a tendency to repel each other
but they can't, being blocked by their neighbors in k-space. The hypothesis
of a complete non-perturbative reorganization of the ground state is
therefore not completely unplausible.
In one dimension the repulsion of particles from common points in momentum
space affects only excitations, not the ground state, which is instead
modified by interactions between opposite Fermi points. However, in two
dimensions the situation may be different, since new gapless degrees of
freedom parallel to the Fermi surface appear.
\par
\pp In summary, in our opinion there are interesting hints but no evidence
for a breakdown of Fermi liquid theory at weak coupling in two dimensions.
Clear is only that such a breakdown would have to be much more subtle that
in one dimension.
\par

\vskip 1cm

{\bf Acknowledgements:}
We would like to thank Phil Anderson, Carlo Di Castro and Michele Fabrizio for
numerous valuable discussions.
This work has been supported by the European Economic Community under Contract
No. SC1* 0222-C(EDB). One of us (W.M.) also gratefully acknowledges the kind
hospitality of the Condensed Matter Group at Princeton University, and
financial support by the Deutsche Forschungsgemeinschaft.

\vfill \eject

{\bf Appendix A: Phase shift, phase angle and overlap in 1D} \par
\medskip
\pp In one dimension, the non-interacting two-particle spectrum
$\lbrace E^0_{\alpha}\rbrace$ of states with fixed total momentum $\bP$
is locally invariant, i.e. the levels become
equidistant in the limit $L \to \infty$, $E^0_{\alpha} \to E$, with a spacing
$\Delta E^0_{\alpha} = E^0_{\alpha+1} - E^0_{\alpha}$ that depends only on
$E$. The level degeneracy $d^0_{\alpha}$ also becomes a function of $E$ only.
In this situation, the phase shift, energy shift and wave function overlap
are uniquely determined by the phase angle, for almost all states in the
large system limit, as we will now show.
\par
\pp For large $L$, and assuming local spectral invariance, the eigenvalue
equation (3.6) can be written as \cite{STA,DEW}
$$ {1 \over U} = K^{loc}_L(E) + \Re K(E)  \eqno(A.1) $$
where $K(E) := \lim_{\eta \to 0} \lim_{L \to \infty} K_L(E+i\eta)$, and
$$ K^{loc}_L(E) = L^{-d} \sum_{m = -\infty}^{\infty}
   {d^0 \over \delta E - m \Delta E^0}  \eqno(A.2) $$
is the "local sum", which is determined by levels in an infinitesimal interval
centered at $E$. Here $\Delta E^0 = \Delta E^0(E)$ is the level spacing, and
$d^0 = d^0(E)$ the degeneracy of levels with energy $E$.
This decomposition holds asymptotically for almost all
levels, exceptions being levels which are so close to the bottom or top
of the spectrum that their local sum does not extend over a large number
of levels on both sides.
Using the identity $\sum_{m = -\infty}^{\infty} (x-m)^{-1}
= \pi\cot(\pi x)$, one obtains
$$ K^{loc}_L(E) = \pi \rho(E) \cot(-\chi(E))  \eqno(A.3) $$
where $\rho(E) = d^0/L^d \Delta E^0$ is the density of states, and
$$ \chi(E) := -\pi \delta E/\Delta E^0  \eqno(A.4) $$
the phase shift. Inserting (A.3) into the eigenvalue equation, and solving
for $\chi(E)$, one finds
$$ \chi(E) = -\tan^{-1}\left[ { U \pi \rho(E) \over 1 - U \Re K(E)} \right]
   \eqno(A.5) $$
Since $\pi \rho(E) = -\Im K(E)$, this is nothing but the phase angle
$\phi(E)$, defined in (3.23). Hence, for locally invariant spectra, phase
shift and phase angle are indeed equivalent for large systems. A slightly
different derivation of this result can be found in Ref. \cite{ER}.
\par
\pp The normalization constant $C$ is given by
$$ C^{-2} = (C^{-2})^{loc} = \sum_{m = -\infty}^{\infty}
   {d^0 \over [\delta E - m \Delta E^0]^2}  \eqno(A.6) $$
for $L \to \infty$ and local spectral invariance. Note that here only levels
in an infinitesimal neighborhood around $E$ contribute. Using the identity
$\sum_{m = -\infty}^{\infty} (x-m)^{-2} = \pi^2/(\sin(\pi x))^2$, one
obtains

$$ C(E) = \sqrt{d^0} {|\sin\chi| \over L^d \pi \rho} \eqno(A.7) $$
The overlap between interacting and non-interacting wave functions is thus
$$ \bra\Phi_{\alpha}|\Psi_{\alpha}\ket =: S(E) = \sqrt{d^0} C/\delta E =
   {\sin\chi(E) \over \chi(E)} \eqno(A.8) $$
Recall that $|\Phi_{\alpha}\ket$ is a symmetric combination of $d^0$
degenerate states with amplitude $1/\sqrt{d^0}$ for each. The overlap
of $|\Phi_{\bk\bk'}\ket$ and $|\Psi_{\bk\bk'}\ket$ is obtained from this
by inserting $E = \eps_{\bk} + \eps_{\bk'}$.
\par

\vfill\eject

%
%

\vfill \eject

%
%
\leftline{\Large\bf Figure Captions}
\begin{enumerate}

\item[{\bf Fig. 1}] Fermi surfaces $\partial\cF^-$ and $\partial\cF^+$
 on the discrete k-lattice in a finite system; the continuous line
 represents the Fermi surface $\partial\cF$ in the large system limit.

\item[{\bf Fig. 2}] Regimes for the two-particle density of states in the
 $(P,\omega)$-plane, separated by the functions $\omega_+(P)$,
 $\omega_-(P)$ and $\omega_0(P)$. The dotted line indicates values for
 $(P,\omega)$ with $\rho = x/4\pi$, $x = 0.3$.

\item[{\bf Fig. 3}] Geometry of available two-particle states with fixed
 total momentum $\bP$ and fixed energy in two dimensions. The bold
 sections on the circle around $\bP/2$ indicate the possible locations
 of momenta $\bk$ and $\bk'$ outside $\cF$ such that $\bk+\bk' = \bP$
 and $\eps_{\bk} + \eps_{\bk'} = {\rm const}$.

\item[{\bf Fig. 4}] Two particles in $\bk' \maceq \bk$ on the same Fermi
 point in 2D: generic case (a) and the two exceptional cases (b) and (c).
 Note that the plot shows an infinitesimal fraction of the Fermi surface,
 which therefore looks perfectly flat.

\item[{\bf Fig. 5}] Phase shifts (points) and phase angle (line) as
 function of energy for a Hubbard model with next-neighbor hopping $t=1$,
 coupling $U = 5$ and $\eps_F = -1.3$ in 1D, for a fixed total momentum
 $\bP = 0$. The phase shifts have been calculated for a finite system
 with $L = 1000$.

\item[{\bf Fig. 6}] Phase shifts (dots), phase angle (solid line) and
 mean phase shift (dashed line) as function of energy for a Hubbard model
 with next-neighbor hopping $t=1$, coupling $U = 5$ and $\eps_F = -2.3$
 in 2D, for a fixed total momentum $\bP = (0.28\pi,0.36\pi)$.
 The phase shifts have been calculated for a finite system with $L = 100$.
 Each point $\bar\chi(E)$ is obtained by averaging 50 phase shifts
 corresponding to eigenenergies near $E$ (the dashed line connects these
 points as guide to the eye).

\item[{\bf Fig. 7}] Phase angle for two particles with momenta $\bk$ and
 $\bk'$ close to a common point of the Fermi surface, plotted for various
 fixed total momenta $\bP = \bk+\bk'$ as a function of the ratio
 $|\bk'-\bk|/(P-2k_F)$. A quadratic $\eps_{\bk}$ with a cutoff
 $\omega_c = 10$, and a constant coupling $U = 5$ has been used in this
 plot.

\end{enumerate}

\end{document}